# Thirty Femtograms Detection of Iron in Mammalian Cells

A. Galimard[1], M. Safi[1], N. Ould-Moussa[1], D. Montero[2], H. Conjeaud[1] and J.-F. Berret[1]*

[1] *Matière et Systèmes Complexes, UMR 7057 CNRS Université Denis Diderot Paris-VII, Bâtiment Condorcet 10 rue Alice Domon et Léonie Duquet, 75205 Paris (France)*
[2] *Institut Des Matériaux de Paris Centre (FR CNRS 2482), Université Pierre et Marie Curie, 75252 PARIS Cedex 05, France*

**Abstract :** Inorganic nanomaterials and particles with enhanced optical, mechanical or magnetic attributes are currently being developed for a wide range of applications. Safety issues have been formulated however concerning their potential cyto- and genotoxicity. For *in vivo* and *in vitro* experimentations, recent developments have heightened the need of simple and facile methods to measure the amount of nanoparticles taken up by cells or tissues. In this work, we present a rapid and highly sensitive method for quantifying the uptake of iron oxide nanoparticles in mammalian cells. Our approach exploits the digestion of incubated cells with concentrated hydrochloric acid reactant and a colorimetric based UV-Visible absorption technique. The technique allows the detection of iron in cells over 4 decades in masses, from 0.03 to 300 picograms per cell. Applied on particles of different surface chemistry and sizes, the protocol demonstrates that the coating is the key parameter in the nanoparticle/cell interactions. The data are corroborated by scanning and transmission electron microscopy and stress the importance of resiliently adsorbed nanoparticles at the plasma membrane.

## 1 - Introduction

Engineered nanoparticles (NPs) have attracted much attention during the last decade because of their size-related properties, large surface-to-volume ratios and high surface reactivity. Inorganic NPs are already applied in biotechnology, nanomedicine and materials science. Safety issues have been formulated however concerning their potential cyto- and genotoxicity. Nowadays, extensive research efforts are directed towards the risk assessment exhibited by nanomaterials. Parameters such as the size, charge and surface chemistry are examined because they affect the interactions of the NPs with living cells. *In vitro* toxicity assays consist in incubating cell or bacteria cultures with NPs and to monitor biomarkers of the cell viability and activity. In recent years, numerous mammalian cell lines were confronted to a wide variety of nanomaterials (for recent reviews, see Refs. [1-3]). In combination with these assays, it is also crucial to determine the amount of NPs actually taken up by the cells. Relevant for the evaluation of the toxicity, this knowledge is also important for applications including drug delivery, diagnostic and cell labeling.

Conventional methods for measuring the amount of internalized nanomaterials are based on coupled plasma spectroscopies, including mass spectrometry (ICP-MS),[4-6] optical emission spectroscopy (ICP-OES)[7] and atomic emission spectroscopy (ICP-AES).[8] These analytical techniques are rapid and accurate, but cumbersome and limited to some elemental species. They also imply the ionization of mineralized samples by a plasma source. For the quantification of various types of nano- and micro-particles in mammalian cells, the innovative techniques of laser desorption/ionization[9] and cell[10] mass spectrometry were proposed, and represent advanced alternatives. More recently, quantitative flow cytometry was also shown to be a straightforward method,[11] from which the proportions of adsorbed *versus* internalized NPs can be determined.[12]

In the present communication, we report on a facile and cost-effective method to quantify the total iron content of mammalian cells labeled with maghemite ($\gamma$-$Fe_2O_3$) NPs. For iron oxide-based materials, specific analytical techniques were designed during the last years, including magnetophoresis,[13, 14] superconducting quantum interference device[15], magnetic resonance imaging,[16, 17] relaxometry[16-18] and UV-visible spectrometry.[18-20] Concerning this later technique, reports from the literature exploit colorimetric assays but mention different chemical reactants for the detection of iron. Combined with the acid digestion of cells, ferrozine, a chelating agent of ferrous ions ($Fe^{2+}$) is probably the most commonly used biomarkers. The UV-visible absorbance of Fe-ferrozine complexes exhibits a characteristic peak at 562 nm and its amplitude scales linearly with concentration. This analytical technique provides a high sensitive detection method,[19] but is not directly applicable to maghemite which contains iron atoms in their Fe(III)-oxidized state. Hydrochloric acid has also been examined as a biomarker for iron, alone,[18, 20] or in conjunction with other acids and reducing agents (*i.e.* potassium ferrocyanide as in the Prussian blue staining protocol).[21] Concerning the sensitivity of the techniques mentioned previously, 1 pg/cell appears as a typical limit.[13-15, 17, 18]

In this work, we develop a new protocol, dubbed MILC for *Mass of metal Internalized/Adsorbed by Living Cells,* which appears as a simple, rapid and highly sensitive method for quantifying the uptake of iron oxide NPs in mammalian cells. Our approach exploits the digestion of incubated cells with concentrated hydrochloric acid reactant and a colorimetric based UV-Vis absorption technique, as described above. In contrast to previous reports, MILC uses the entire absorption spectrum between 200 nm and 1000 nm and the fitting against calibrated references to derive the iron content. As a result, this new technique allows the detection of iron in cells over 4 decades in masses, typically between 0.03 to 300 pg/cell. Applied on maghemite NPs of different coating and sizes, MILC demonstrates that the coating is the key parameter in the NP/cell interactions. The data are corroborated by scanning and transmission electron microscopy, and stress the importance of resilient adsorbed NPs at the cell membrane.

# 2 - Results and discussion

## 2.1 - The MILC protocol

The amount of iron taken up by the cells was measured following the protocol MILC. MILC makes use of UV-Vis spectrometry to determine the iron concentration from pelleted cells dissolved in concentrated hydrochloric acid.

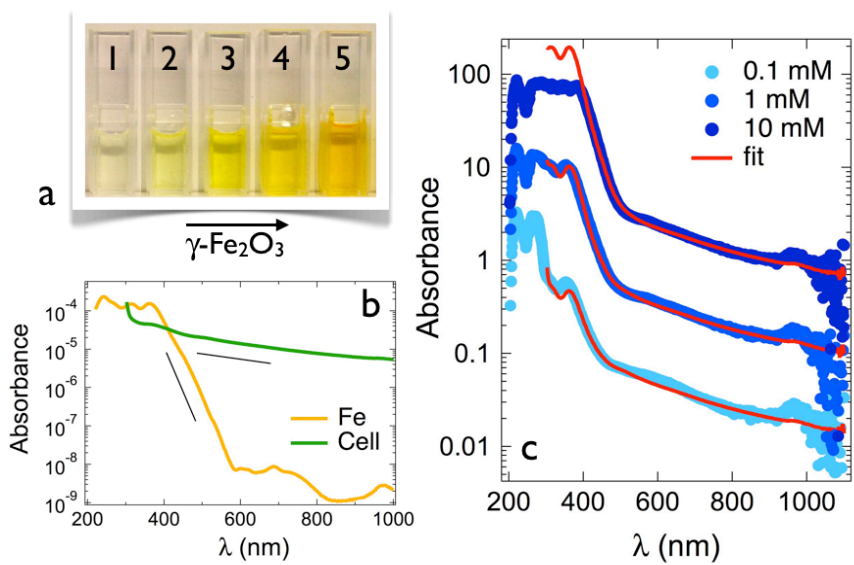

*Figure 1*: a) Iron oxide nanoparticle dissolved in hydrochloric acid (35 vol. %) at various concentrations: (1) [Fe] = 0.1 mM; (2) [Fe] = 0.3 mM; (3) [Fe] = 1 mM; (4) [Fe] = 3 mM; (5) [Fe] = 10 mM. In this work, the concentrations are defined as the iron molar concentration [Fe] or as the percentage by weight of $\gamma$-$Fe_2O_3$. [Fe] = 1 mM corresponds to c($\gamma$-$Fe_2O_3$) = 8×$10^{-3}$ wt. % or 80 μg/ml. b) Absorbance of HCl-based solutions containing respectively 1 pg/μl of iron (orange) and 1 cell/μl (green). The straight lines indicate exponential decreases of the absorbance versus wavelength. c) Absorbance of centrifuged pellets dissolved in HCl at different $\gamma$-$Fe_2O_3$ concentrations, [Fe] = 0.1 (light blue), 1 (blue) and 10 mM (dark blue). Experiments were performed on 3×$10^6$ cells and incubation time of 24 h. The continuous lines through the data points are linear combinations of the two reference curves shown in b). The ratio between the coefficients is $M_{Fe}$.

The results are the mass of iron, $M_{Fe}$ expressed in picograms per cell. We provide here a description of the data analysis, and refer to the Materials and Methods Section for the description of the methodology. MILC proceeds in three steps. The live cells are first cultured and incubated with the particles for times between $t_{Inc}$ = 5 mn to 24 h. The supernatant is removed and the cells are thoroughly washed with PBS. The cells are then trypsinized, numbered using a Mallasez chamber and centrifuged to get a pellet at the bottom of the Falcon© tube. The pellets are dissolved in hydrochloric acid (35 %), and later investigated by UV-Vis spectrophotometry. The cell pellets dissolved in HCl display the yellow color characteristic of tetrachloroferrate ions $FeCl_4^-$ (Fig. 1a). The absorbance of the dissolved pellets is compared to those of iron oxide and of fibroblasts determined separately. Fig. 1b shows the absorbance curves at respective concentrations 1 pg/µL for γ-Fe$_2$O$_3$ and at 1 cell/µL for NIH/3T3. Over broad ranges of wavelengths, both exhibit characteristic dependences such as exponential decays of the form Abs($\lambda$) ∼ exp (−$\lambda/\lambda_0$). Exponential decreases indicated by straight lines in the figure were obtained for $\lambda_0$ = 21 nm and 358 nm, respectively. For fibroblasts treated with NPs, the absorbance expresses as a linear combination of the two above contributions. The number densities of cells retrieved from the absorbance were checked with direct Malassez counting for further validation. Fig. 1c displays the UV-Vis absorbance of a series of cell samples prepared with increasing iron oxide concentrations, [Fe] = 0.1, 1 and 10 mM. The continuous lines through the data points are linear combinations of the two reference curves shown in Fig. 1b, from which the masses of iron per cell are derived. One gets here $M_{Fe}$ = 1, 4 and 12 pg/cell respectively, with an uncertainty of 10%. Assuming for the absorbance an absolute uncertainty of 0.03, the minimum amount of iron detectable by this technique is 0.03 pg/cell, *i.e.* 30 fg/cell. With MILC, the high sensitivity arises from the fact that the whole UV-Vis spectrum is taken into account in the adjustment.[18-20]

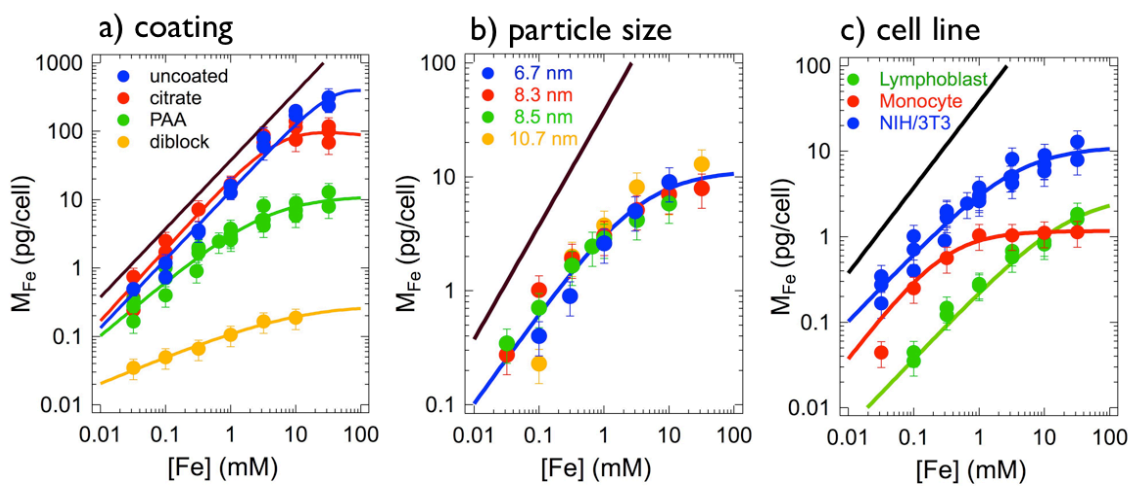

*Figure 2* : Masses $M_{Fe}$ of internalized and adsorbed iron obtained by varying different parameters, including a) the coating, b) the particle size and c) the cell type. In a), the legends "uncoated", "citrate", "PAA" and "diblock" refer to bare nanoparticles, to Cit–γ-Fe$_2$O$_3$, PAA$_{2K}$–γ-Fe$_2$O$_3$ and PAA$_{5K}$-b-PAM$_{30K}$–γ-Fe$_2$O$_3$, respectively. For the data in a) and c), the particles had a median size of 8.3 nm. For PAA$_{2K}$–γ-Fe$_2$O$_3$, the same data are displayed in the three charts and correspond to NPs sizes 6.7, 8.3, 8.5 and 10.7 nm. The straight lines show the maximum iron content available per cell for a given concentration [Fe] in the supernatant. Experiments were performed on 3×10$^6$ cells and incubation time of 24 h. The continuous color lines are guides for the eyes only.

2.2 – Effect of coating, particle size and cell line

Fig. 2 displays the masses of internalized and adsorbed iron obtained by varying different physico-chemical and biological parameters: the coating of the particles (Fig. 2a), the particle size (Fig. 2b) and the cell type (Fig. 2c). The straight line in Figs. 2 depicts the maximum amount of iron $M_{Fe}^{Max}$ that can be taken up by a cell. It is calculated by dividing the mass of iron present in the supernatant by the number of cells in the assay ($3\times10^6$ fibroblasts exposed to a concentration of 1 mM correspond to $M_{Fe}^{Max}$ = 37 pg/cell). Note that this limit is independent on the particle size or aggregation state, and thus appears similarly in the three diagrams. For uncoated and citrate coated NPs, $M_{Fe}$ increases linearly with [Fe] and levels off above 10 mM. Here, the masses of internalized/adsorbed iron are high and above 100 pg/cell. In the linear parts, *i.e.* below 3 mM, it represents about 30% of the maximum value $M_{Fe}^{Max}$. For neutral PAA$_{5K}$-*b*-PAM$_{30K}$ and charged PAA$_{2K}$ coating, the [Fe]-variations are similar but the saturation plateaus are found at much lower levels (0.2 and 10 pg/cell as compared to 100 pg/cell for citrate). As shown in Fig. 2a, MILC allows the detection of iron in cells over 4 decades in masses, typically between 0.03 to 300 pg/cell. Fig. 2b explores the role of the NP size on the internalized amount, all NPs being now coated with PAA$_{2K}$. The mass of iron per NIH/3T3 cell is displayed for magnetic cores varying between 6.7 and 10.7 nm (Supporting Information SI-1). The data of the four specimens agree well with each other, suggesting that for this specific coating, there is no major effect of the NP size on the internalization amounts. Fig. 2c compares the $M_{Fe}$-values for cells adhering on a substrate (NIH/3T3 fibroblast) and in suspensions (2139 lymphoblasts, THP1 monocytes). PAA$_{2K}$–$\gamma$-Fe$_2$O$_3$ NPs with core size 8.3 nm were used in this assay. Lymphoblastoid and monocytes are in the 0.1 – 1 pg/cell range that is slightly lower than for fibroblasts. From these experiments, it is found that the coating is the most important parameter affecting the internalized/adsorbed amounts of iron,[22, 23] with amplitudes of variation larger than 1000 between the saturation levels (0.2 pg/cell for PAA$_{5K}$-*b*-PAM$_{30K}$ against 300 pg/cell for uncoated NPs). Cell type plays a minor role, whereas NP size has no noticeable impact. This later result could be explained by the fact that the sizes of the PAA$_{2K}$-coated particles cover a narrow range in the present work. The diameters of the magnetic cores range for instance from 6.7 nm to 10.7 nm (SI-1). Previous studies have reported size dependent uptake, but for NPs spanning over a broader size range, typically up to 1 μm.[23-26] Although these studies did not discuss the stability of the NPs in biological fluids, it is found as a general trend that the numbers on NPs internalized decrease with the size.

2.3 – Localization of particles inside the cells

Fibroblasts seeded with iron oxide NPs were further investigated by TEM. Fig. 3a and 3d provides representative images of NIH/3T3 cells incubated with Cit–$\gamma$-Fe$_2$O$_3$ and PAA$_{2K}$–$\gamma$-Fe$_2$O$_3$, respectively. The experimental conditions were an incubation time of 24 h and an iron concentration of 10 mM. The corresponding $M_{Fe}$-levels were 100 ± 20 and 7 ± 2 pg/cell respectively (Fig. 2a). A careful analysis of the TEM images shows that the NPs were primarily located in membrane-bound compartments, or endosomes.[13, 14, 27-30] Close-up views of the selected areas (rectangles) clearly identify the lipidic membrane separating the cytosol from the NPs (Fig. 3b and 3e). In this work, NPs were found neither in the cytosol nor in the nucleus. A statistical analysis of the endosome sizes was performed and revealed weak variations as a function of concentration and coating. For citrate coated NPs at 1 and 10 mM, the endosomal distribution were peaked at 500 and 800 nm respectively (Fig. 3c). Compartments larger than 1 µm were also detected. For the $PAA_{2K}$–γ-$Fe_2O_3$ at 10 mM, the average size was 500 nm, *i.e.* in relative accordance with those of the cell control (data not shown). The similarities of the iron oxide loaded compartments in Figs. 3a and 3d suggest similar mechanisms of entry into the cells, *i.e.* endocytosis.[3, 31] The major difference between the two coating specimen lies in the spatial distribution of the particles inside endosomes: with citrate, the NPs appear as aggregated under the form of clusters, whereas with polymers they are randomly spread and unassociated.

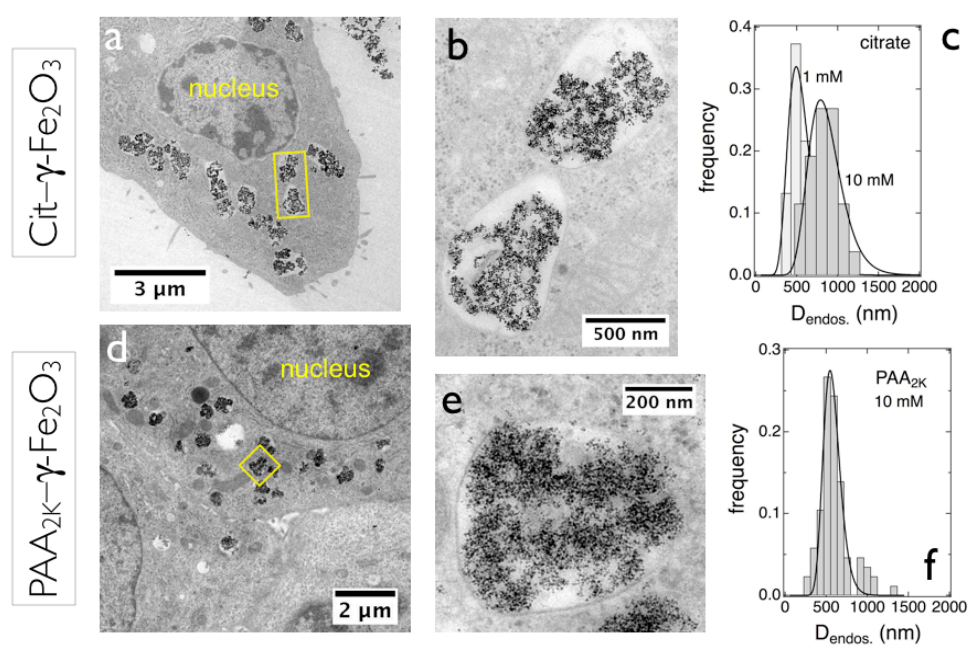

*Figure 3:* Transmission electron microscopy images of NIH/3T3 fibroblast cells incubated wit: a) Cit–γ-$Fe_2O_3$ and with d) $PAA_{2K}$–γ-$Fe_2O_3$ (with $t_{Inc}$ = 24 h and [Fe] = 10 mM). b) and e) are close views of the delimited areas in a) and d). c) and f) size distribution of the endosomes at concentrations 1 mM and 10 mM. With the units used in this work, [Fe] = 1 mM corresponds to c(γ-$Fe_2O_3$) = 8×$10^{-3}$ wt. % or 80 µg/ml.

With the $PAA_{2K}$-coated NPs, the endosomes were also more homogeneously filled. In conclusion to this part, and in good agreement with the literature,[12, 13, 27-30, 32] we have found that the carboxylate coated NPs are internalized by the NIH/3T3 fibroblasts and located in endosomal membrane-bound compartments. The coating has here a moderate impact on the endosome size distributions. These results contrast with those obtained by the MILC technique that highlights large differences in the internalized/adsorbed iron masses. They also suggest that large quantities of NPs should be extracellular and located on the plasma membrane.

## 2.4 – Adsorption and internalization kinetics

Fig. 4 displays the variations of the mass of internalized and/or adsorbed iron with the incubation time $t_{Inc}$ upon addition of uncoated, citrate and PAA$_{2K}$-coated particles. The investigated concentrations were [Fe] = 4 mM for the bare particles, and 1 and 10 mM for the coated specimens. As noted earlier, the evolution of the recorded masses at 37 °C imparts two interaction mechanisms: the adsorption of the NPs on the plasma membrane and their internalization into endocytic vesicles.[12, 13, 28, 29]

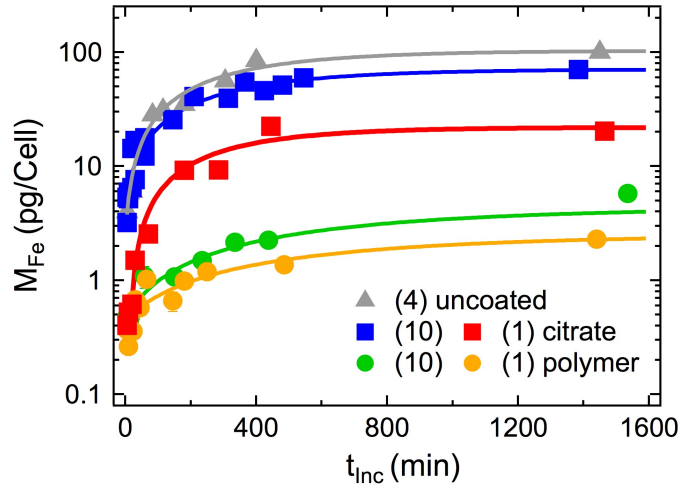

*Figure 4 : Incubation time dependence of the internalized and/or adsorbed iron content in NIH/3T3 cells upon addition of uncoated, citrate and PAA$_{2K}$-coated particles. Concentrations in mM are indicated in parenthesis. Note that for uncoated and citrate coated NPs at 4 and 10 mM respectively, the mass of iron per cell exhibits an initial jump within the first 5 minutes of contact with NPs. The continuous lines are exponential growth functions given by Eq. 1.*

For the present case studies, the mass of internalized/adsorbed iron increases following a similar pattern: within the first 5 minutes of contact with NPs, $M_{Fe}(t_{Inc})$ jumps to an initial value noted $M_{Fe}^0$ which depends on both concentration and coating. This initial jump is followed by a progressive leveling-off towards a saturation value noted $M_{Fe}^\infty$. In these assays, the $M_{Fe}^\infty$-values were consistent with those recorded at 24 h (Fig. 2a). The continuous lines between the data points are exponential growth functions of the form:

$$M_{Fe}(t_{Inc}) = M_{Fe}^0 + \Delta M_{Fe}\left(1 - exp\left(-\frac{t_{Inc}}{\tau}\right)\right) \qquad (1)$$

where $\Delta M_{Fe} = M_{Fe}^\infty - M_{Fe}^0$ denotes the difference between the final and initial states and $\tau$ the characteristic time of the kinetics. For the time profiles in Fig. 4, the $\tau$'s are of the order of 300 - 500 min for all specimens tested (Table I). Similar exponential profiles together with a leveling-off of the uptake were found in the recent literature.[10, 12-14, 21] Interestingly, these behaviors did not depend on the cell type. Note that for PAA$_{2K}$-coated NPs at 10 mM the measured mass at 5 min is much lower than that of the citrate or uncoated particles (0.44 versus 3.2 pg/cell). This later result suggests that in case of an initial $M_{Fe}$-jump, a large quantity of nanoparticles should be adsorbed at the surface of the cells. Some studies have also reported similar trends.[10, 13, 21]

| $\gamma$-Fe$_2$O$_3$ | [Fe] (mM) | $M_{Fe}^0$ (pg/cell) | $\Delta M_{Fe}$ (pg/cell) | $\tau$ (min) |
|---|---|---|---|---|
| uncoated | 4 | 4 | 104 | 350±50 |
| citrate-coated | 1 | 0.4 | 22 | 330±100 |
|  | 10 | 3.2 | 70 | 320±40 |
| PAA$_{2K}$-coated | 1 | 0.36 | 2.4 | 530±100 |
|  | 10 | 0.44 | 4.0 | 490±140 |

**Table I**: *List of the fitting parameters obtained from the adjustment of the adsorption and internalization kinetics data (Eq. 1). Here, $M_{Fe}^0$ is the mass of iron extrapolated at $t_{Inc} = 0$ and $\Delta M_{Fe}$ the difference $M_{Fe}^\infty - M_{Fe}^0$, where $M_{Fe}^\infty$ is the mass of iron extrapolated at $t_{Inc} = \infty$. $\tau$ denotes the relaxation time of the process of the exponential rise.*

## 2.5 – Particles at the plasma membrane

Scanning Electron Microscopy (SEM) experiments were performed to visualize the plasma membrane of cells treated with Cit–$\gamma$-Fe$_2$O$_3$ and PAA$_{2K}$–$\gamma$-Fe$_2$O$_3$ NPs. The experimental conditions investigated were an incubation time of 2 h and an iron concentration of 10 mM. The corresponding $M_{Fe}$-levels found by MILC were 24 ± 5 pg/cell and 1.1 ± 0.2 pg/cell respectively, *i.e.* ~ 20 times larger for citrate than for PAA$_{2K}$ (Fig. 4).

Fig. 5a shows a SEM image of a control untreated cell. The cylindrical body of the NIH/3T3 exhibits at its surface numerous protrusions identified as microvilli and indicated by arrows.[26, 28, 33-37] Formed as cell extensions starting from the membrane, microvilli are involved in a wide variety of functions, including internalization, cellular adhesion and mechano-transduction. For NIH/3T3 fibroblasts, their average length, diameter and density are respectively 600 nm, 160 nm and 3 μm$^{-2}$, in good agreement with literature data[38] (see Supporting information SI-2).

Fig. 5b and 5c display representative images of the plasma membrane for a cell incubated with PAA$_{2K}$–$\gamma$-Fe$_2$O$_3$ NPs. The cell body and the microvilli covering the cell surface maintained their morphologies after a 2 h exposure. The average length and density of the protrusions remained also unchanged at 690 nm and 3 μm$^{-2}$ respectively (SI-2). Scattered on the cell surface, the NPs were difficult to localize. On a few locations, SEM showed particles attached to the cell membrane as single entities or as clusters. In some other cases, the microvilli appeared as entangled by the clusters,[26, 28, 37] as identified in Fig. 5c or acquired a more spherical shape (size 200 nm).[34] Fibroblasts incubated with citrate-coated particles appeared very differently. Fig. 5d provides an image of a cell that is merely covered with particles and particle clusters. There, the deposited materials form a thick and porous layer of aggregates, the largest being of a few microns (arrowhead). On this image the plasma membrane cannot be seen. Citrate-coated particles also induced a dramatic transition in the morphology of the microvilli. They are now elongated threads with average length 1.2 μm, going up to 3 μm. In a few examples, the microvilli are terminated with a sphere covered with an iron oxide cap (inset). Other SEM images obtained with citrate show similar features, but with less adsorbed material at the membrane. Phase-contrast optical microscopy evidenced that NP clusters were still present at the cell surface after an incubation of 24 h (SI-3), confirming the strong resilience of the adsorbed layer. In conclusion, SEM studies demonstrate the importance of visualization for distribution of particles on a cellular and subcellular level. This technique allows inferring the large $M_{Fe}$-values found by MILC for Cit–γ-Fe$_2$O$_3$ to extracellular iron oxide adsorbed at the plasma membrane in the form of highly heterogeneous clusters.

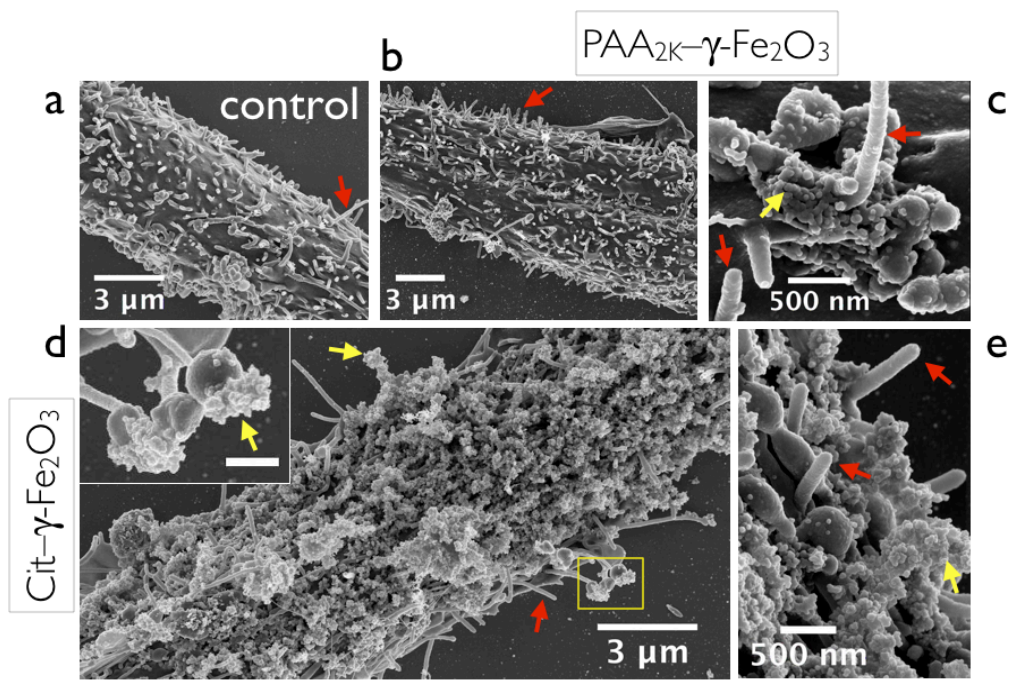

*Figure 5*: Scanning Electron Microscopy images of NIH/3T3 untreated cells (a) and of cells treated with $PAA_{2K}$–γ-Fe$_2$O$_3$ (b,c) and with Cit–γ-Fe$_2$O$_3$ (d,e). Incubation time was 2 h and the concentration in the supernatant 10 mM. The protrusions at the cell surface were identified as microvilli (red arrows).[26, 28, 35-37] With the $PAA_{2K}$ coating, few NPs appeared at the membrane, whereas with citrate, the cell is covered with a 500 nm thick and porous layer of aggregates. Microvilli are then longer as compared to the control (inset). Adsorbed particles are marked with yellow arrows. Additional visualization of the cell surfaces proved that in the case of citrate the NP layer was resiliently adsorbed, meaning that it did not desorb nor it was internalized.

2.6 – Adsorption at the cell membranes and colloidal instability

The bare and the citrate-coated NPs exhibit the highest masses of adsorbed/internalized iron. In culture media such as RPMI and DMEM, the particles aggregate and form highly polydisperse and micronic clusters[39, 40]. For Cit–γ-Fe$_2$O$_3$, this destabilization was studied recently and was attributed to the displacement of the citrates from the particle surfaces towards the bulk, as they are preferentially complexed by calcium and magnesium counterions of the culture medium. For non-coated particles, the precipitation occurred because of the change of pH between that of the synthesis medium (pH 2) and that of the culture medium (pH 7.4). To study the kinetics of aggregation, light scattering experiment was performed using the same protocol as for the incubation *in vitro*. In these assays, a few microliters of a concentrated iron oxide dispersion was poured rapidly in a test tube containing the cell culture medium, the hydrodynamic diameter being thereafter monitored as a function of the time (SI-4). For Cit–γ-Fe$_2$O$_3$, the hydrodynamic diameter was shown to increase within the first seconds of contact from 23 to 180 nm and remained at this level for one hour. For uncoated NPs, the initial increase was even more important ($D_H > 1\ \mu m$). At longer times, the aggregates settled down at the bottom of the tube, resulting in a decrease of the scattering intensity and hydrodynamic diameter. Similar observations were made in Petri dishes when the particles were incubated with cells. At this point, it is unclear whether some NPs precipitate directly on the cell membranes, or if there is first the formation of aggregates and then a deposition of the cells. However, the similarity of behaviors for $M_{Fe}$ (Fig. 4) and $D_H$ (SI-4) at short times supports the first scenario. It is also plausible that the cells act as sites of nucleation for the destabilization process. On a longer time scale, the sedimentation of the NP clusters down to the adherent cell layer enters into play and further increases the adsorption of the NPs, in agreement with several earlier reports.[4, 41] From the above data, we conclude that the strong NP adsorption seen by SEM and optical microscopy is heightened by the destabilization of the NPs in the cell culture media.

# 3 - Conclusion

In the present paper we propose a simple and facile assay to measure the iron content of living mammalians cells incubated with iron oxide NPs. This assay is based on UV-Vis spectrometry and on the digestion of seeded cells with hydrochloric acid. With a 10%-accuracy and a minimum sensitivity as low as 30 femtograms per cell, the technique allows the dosage of iron in cells over 4 decades in masses, measuring the iron content and cell number at once. The present study was also designed to examine the effect of physico-chemical parameters on the NP/cell interactions.

One of the more significant findings to emerge from this work is the correlation between the phase behavior of the NPs in biological fluids and their interactions with adherent mammalian cells. Iron oxide NPs coated with citrate ligands are found to destabilize in cell culture media and to adsorb massively onto NIH/3T3 fibroblast cells. Particles made from the same magnetic core, but coated with a 3 nm polymer adlayer exhibit much weaker interactions with cells. These findings confirm those obtained with human lymphoblasts in suspension and treated with the same particles.[12] The most striking result revealed here came from scanning electron microscopy, which disclosed huge amounts of resiliently adsorbed NPs at the cell membrane. This technique permits to assign the large iron contents found for Cit–$\gamma$-$Fe_2O_3$ (typically 100 pg/cell at high [Fe]) to extracellular iron oxides adsorbed at the plasma membrane in the form of highly heterogeneous clusters. In parallel, the low amount of $PAA_{2K}$–$\gamma$-$Fe_2O_3$ NPs detected by SEM at the cell membrane suggests that the MILC-data for this compound (around 10 pg/cell at saturation) corresponds mainly to the iron content internalized by the cells. Such data are in good agreement with those of the literature.[10, 15, 20, 21] Given the wide use of citrates for coating engineered NPs (as in the cases of gold[42-44] and rare-earth oxide,[45] the present results, together with those of the recent literature[46-48] confirm the poor coating performances of such low-molecular weight ligands. Taken together, these results suggest two types of NP-cell behaviors: NPs are either adsorbed on the cell membranes, or internalized into membrane-bound endocytic compartments. In the case of adsorption, as shown by SEM, the layer of materials deposited on the membranes can grow up to 500 nm and is still observed after 24 h of incubation. More importantly, the present method for the measurement of internalized/adsorbed iron in living environments can be easily extended to other living environments such as bacteria, tissues and organs.

# 4 - Materials and Methods

Chemicals, synthesis and characterization

*Nanoparticles synthesis*: the iron oxide nanoparticles (bulk mass density $\rho$ = 5100 kg m$^{-3}$) were synthesized according to the Massart method[49] by alkaline co-precipitation of iron(II) and iron(III) salts and oxidation of the magnetite ($Fe_3O_4$) into maghemite ($\gamma$-$Fe_2O_3$). The nanoparticles were then size-sorted by subsequent phase separations.[50] At pH 1.8, the particles are positively charged, with nitrate counterions adsorbed on their surfaces. The resulting interparticle interactions are repulsive, and impart an excellent colloidal stability to the dispersion. For the present study, three batches of $\gamma$-$Fe_2O_3$ NPs of median diameter 6.7, 8.3 and 10.7 nm were synthesized. The size and size distribution were retrieved from Vibrating Sample Magnetometry (VSM) and from Transmission Electron Microscopy (TEM), and were found in good agreement. Table II lists the results obtained for the median diameters noted $D_0^{VSM}$ and $D_0^{TEM}$ and for the polydispersities. The polydispersity was defined as the ratio between standard deviation and average diameter. VSM was also used to determine the volumetric magnetization $m_S$ of the NPs, which was found to be 2.9×10$^5$ A m$^{-1}$, *i.e.* slightly lower than the volumetric magnetization of bulk maghemite ($m_S$ = 3.9×10$^5$ A m$^{-1}$).

From electron microdiffraction scattering, the crystallinity of the NPs was demonstrated by the appearance of five diffraction rings which wave vectors matched precisely those of the maghemite structure (Supporting Information SI-1). In contrast to recent investigations (see Ref.[12] and reference therein), in the pH range 1.8-10 and for different coating and environments probed, the release rate of ferric ions was found to be negligible (Supporting Information SI-5). In this work, the NP concentrations are defined by the percentage by weight of γ-Fe$_2$O$_3$ in the dispersion or by the iron molar concentration [Fe]. With these units, c(γ-Fe$_2$O$_3$) = 8×10$^{-3}$ wt. % or 80 µg/ml corresponds to [Fe] = 1 mM. For the particles of diameter 8.3 nm, this concentration corresponds to a density of 8.3×10$^{12}$ ml$^{-1}$.

*2.1.2 - Coating*

To improve their colloidal stability and in particular in biologic fluids, different types of electrostatic-based coating were developed. The complexation of the surface charges with ligands such as citrate ions was performed during the particle synthesis. It allowed reversing the surface charge of the NPs from cationic at low pH to anionic at high pH through the ionization of the carboxyl groups. At pH 8, the particles, described as Cit–γ-Fe$_2$O$_3$ in the following were stabilized by electrostatic interactions mediated by the anionically charged ligands. As a ligand, citrate ions were characterized by an adsorption isotherm characterized by a charge density at saturation of 2.14 nm$^{-2}$ and by an affinity constant of 1560 M$^{-1}$. In this context, a 8.3 nm citrate-coated NP bears an average charge of -430$e$. [12, 45, 50]

| Size and polydispersity of γ-Fe$_2$O$_3$ | | | | |
|---|---|---|---|---|
| $D_0^{VSM}$ (nm) | 6.7 | 8.3 | 8.5 | 10.7 |
| $s^{VSM}$ | 0.21 | 0.26 | 0.29 | 0.33 |
| $D_0^{TEM}$ (nm) | 6.8 | 9.3 | 10.7 | 13.2 |
| $s^{TEM}$ | 0.18 | 0.18 | 0.21 | 0.23 |

| coating | hydrodynamic diameter $D_H$ (nm) | | | |
|---|---|---|---|---|
| citrate | 12.7 | 24.4 | 22.9 | 27.0 |
| PAA$_{2K}$ | 17.7 | 30.7 | n.d. | 35.0 |

*Table II*: *Characteristics of the iron oxide particles used in this work. $D_0^{VSM}$ and $D_0^{TEM}$ denote the median diameter of the bare particles determined by Vibrating Sample Magnetometry (VSM) and by Transmission Electron Microscopy (TEM). Similarly, $s^{VSM}$ and $s^{TEM}$ are the values of the polydispersity for the size distributions. $D_H$ is the hydrodynamic diameter of particles dispersed in water and determined by dynamic light scattering. The $D_H$'s for the bare and citrate-coated particles are identical. Values of the hydrodynamic diameters in the cell culture medium are given in SI-4.*[39]

The cationic particles were also coated with $M_W$ = 2000 g mol$^{-1}$ poly(acrylic acid) (PAA$_{2K}$) using the *precipitation-redispersion* process.[51] The polymer of polydisperstiy 1.7 was purchased from Sigma Aldrich and used without further purification. To adsorb the PAA$_{2K}$ chains onto the particle surface, the particles were first precipitated into a dilute solution containing a large excess of polymers (pH 1.8). The precipitate was separated by centrifugation and its pH was increased by addition of ammonium hydroxide. The precipitate redispersed spontaneously at pH 7-8, yielding a clear solution containing the individual polymer coated particles, dubbed PAA$_{2K}$–$\gamma$-Fe$_2$O$_3$ in the following. This process resulted in the adsorption of a highly resilient polymer adlayer surrounding the particles. The hydrodynamic diameters of the bare, citrate- and PAA$_{2K}$-coated particles in DI-water were determined by dynamical light scattering (Table II). From these values, the layer thickness was estimated at 3 ± 1 nm. Hydrodynamic sizes of the particles dispersed in DMEM after one day and after one week are provided in SI-4.[39] The density of chargeable carboxylic groups was evaluated by acid titration at 25 ± 3 nm$^{-2}$. For the particles of diameter 8.3 nm, it corresponded to 5400 structural anionic charges in average. As a final step, the dispersions were all dialyzed against DI-water which pH was first adjusted to 8 (Spectra Por 2 dialysis membrane with MWCO 12 kD). At this pH, 90 % of the carboxylate groups of the citrate and PAA$_{2K}$ coating were ionized.

The synthesis of the copolymer poly(acrylic acid)-*b*-poly(acrylamide) was based on the Madix technology which uses the xanthate as chain-transfer agent in the controlled radical polymerization.[52] The molecular weights targeted by the synthesis were 5000 and 30000 g mol$^{-1}$ for the charged and neutral blocks respectively. The molecular weight of the chain as determined from static light scattering was slightly higher, at $M_W^P$ = 43 500 ± 1000 g mol$^{-1}$. Dynamic light scattering revealed a hydrodynamic diameter $D_H$ = 11 nm and size exclusion chromatography a polydispersity index $M_W/M_n$ = 1.6. The adsorption of the PAA$_{5K}$-*b*-PAM$_{30K}$ layer was realized *via* an entropy-driven ligand exchange process between citrates ions and copolymers. The hydrodynamic diameter of the resulting core-shell colloid was 50 nm.

*Cell culture and cellular growth* : In this work, one type of adherent cells (NIH/3T3 fibroblast) and two types of cells in suspensions (human lymphoblasts 2139 and the THP1 monocytes) were studied. NIH/3T3 fibroblast cells from mice were grown in T25-flasks as a monolayer in Dulbecco's Modified Eagle's Medium (DMEM) with high glucose (4.5 g L$^{-1}$) and stable glutamine (PAA Laboratories GmbH, Austria). This medium was supplemented with 10% fetal bovine serum (FBS) and 1% penicillin/streptomycin (PAA Laboratories GmbH, Austria). Exponentially growing cultures were maintained in a humidified atmosphere of 5% CO$_2$ and 95% air at 37°C, and in these conditions the plating efficiency was 70 – 90% and the cell duplication time was 12 – 14 h. Cell cultures were passaged twice weekly using trypsin–EDTA (PAA Laboratories GmbH, Austria) to detach the cells from their culture flasks and wells. The cells were pelleted by centrifugation at 1200 rpm for 5 min. Supernatants were removed and cell pellets were re-suspended in assay medium and counted using a Malassez counting chamber.

The lymphoblastoid cell line 2139 is a normal line provided to us by Dr Janet Hall from the Institut Curie (Orsay, France). Lymphoblasts are one of the different stages of physiological differentiation inside the lymphoïd line leading to lymphocytes. This cell line was immortalized by the virus Epstein-Barr (EBV) and was obtained by Dr Gilbert Lenoir from the Institut Gustave Roussy (Villejuif, France).[53, 54] The human acute monocytic leukemia cell line (THP1) was provided to us by the Karlsruhe Institute of Technology (Karlsruhe, Germany). Both lymphoblasts and monocytes were grown in suspension in T25-flasks in Roswell Park Memorial Institute (RPMI) with high glucose (2.0 g L$^{-1}$) and stable glutamine. RPMI was supplemented with 10% FBS and 1% penicillin/streptomycin. The culture and counting protocols for the cells in suspension were similar to those of the NIH/3T3 fibroblasts. Note finally that the lymphoblasts and monocytes have a duplication time of 20 h.

Experimental Methods

*MILC protocol*: Cells were seeded onto 3.6 cm Petri dishes, incubated until reaching 60% confluence and then incubated with NPs at different concentrations for times comprised between 5 min and 24 h. The concentrations in the supernatant were varied from [Fe] = 0.03 mM to 30 mM. Note that the concentrations investigated are representative of those reported in the literature for *in vivo* and *in vitro* assays.[3, 55] Previous cytotoxicity studies have shown that γ-Fe$_2$O$_3$ NPs were non-toxic, as the cell viability remained around 100%[40] over periods from 1 to 4 days (SI-6). After the incubation period, the supernatant was removed and the layer of cells washed thoroughly with PBS. The cells were then trypsinized and mixed again with culture medium without serum. 20 µl-aliquots were taken up for counting using a Malassez counting chamber. The cells were finally centrifuged and pellets were dissolved in 35 vol. % HCl. The cells dissolved in HCl were poured in a UV-Vis microcell, studied with a Variant spectrophotometer (Cary 50 Scan) and calibrated with the help of a reference.[17] For the absorbance curve of tetrachloroferrate ions displayed in Fig. 1b, the concentration was determined by flame atomic absorption spectroscopy using a Perkin-Elmer AA100 spectrometer. MILC protocol was performed in triplicate for different types of nanoparticles and coating (for data treatment, see Sections II.2 and SI-7).

*Transmission Electron Microscopy*: TEM on nanomaterials was carried out on a Jeol-100 CX microscope at the SIARE facility of Université Pierre et Marie Curie (Paris 6). It was utilized to characterize the sizes of the γ-Fe$_2$O$_3$ NPs (SI-1). For the TEM studies of cells, the following protocol was applied. NIH/3T3 fibroblast cells were seeded onto the 6-well plate, after the 24 h incubation with NPs, excess medium was removed, and the cells were washed in 0.2 M phosphate buffer (PBS), pH 7.4 and fixed in 2% glutaraldehyde-phosphate buffer 0.1 M for 1 h at room temperature. Fixed cells were washed in 0.2 M PBS. Then, they were postfixed in 1% osmium-phosphate buffer 0.1 M for 45 min at room temperature in dark conditions. After 0.1 M PBS washes, the samples were dehydrated in increasing concentrations of ethanol.

Samples were then infiltrated in 1:1 ethanol:epon resin for 1 h and finally in 100% epon resin for 48 h at 60°C for polymerization. 90 nm-thick sections were cut with an ultramicrotome (LEICA, Ultracut UCT) and picked up on copper-rhodium grids. They were then stained for 7 min in 2% uranyl acetate and for 7 min in 0.2% lead citrate. Grids were analyzed with a transmission electron microscope (ZEISS, EM 912 OMEGA) equipped with a $LaB_6$ filament, at 80 kV and images were captured with a digital camera (SS-CCD, Proscan 1024×1024), and the iTEM software.

*Scanning Electron Microscopy*

Cells were primarily fixed 1 h at room temperature and overnight at 4 °C by immersion in a fixative solution (2.5% glutaraldehyde in 0.1 M sodium cacodylate buffer, pH 7.5), washed 3 times with cacodylate buffer 0.2 M, and post fixed during 1h at room temperature in 1 % osmium tetroxide in 0.1 M sodium cacodylate buffer (pH 7.5) and washed again 3 times in 0.1 M cacodylate buffer. Dehydration until 100 % ethanol was completed through graded ethanol-water mixtures at room temperature. Cells were then dried according to the $CO_2$ critical point drying method (using a Bal-Tec CPD030). After mounting onto scanning stubs samples were coated with a conductive layer (10 nm) of carbon using a thermal carbon evaporator (Cressington C208). Scanning electron microscopy was performed either on a Zeiss ULTRA 40 or a Hitachi SU-70 field emission scanning electron microscopes. Microanalysis and mapping was performed either with Edax CDU or Oxford X-Max EDX detectors installed on the microscopes columns.

# Acknowledgments

We thank Jérôme Fresnais, Jean-Pierre Henry, Olivier Sandre, Emek Seyrek and Michel Seigneuret for fruitful discussions. The Laboratoire Physico-chimie des Electrolytes, Colloïdes et Sciences Analytiques (UMR Université Pierre et Marie Curie-CNRS n° 7612) is acknowledged for providing us with the magnetic nanoparticles. We also thank Frédéric HERBST from the Laboratoire Interdisciplinaire Carnot de Bourgogne for his help in the scanning electron microscopy studies. This research was supported in part by the Agence Nationale de la Recherche under the contract ANR-09-NANO-P200-36 and by the Région Ile-de-France in the DIM framework related to Health, Environment and Toxicology (SEnT).

# Supporting Information

The Supporting Information section SI-1 provides a complete characterization of the iron oxide NPs in terms of structural and magnetic properties. The diameter and length distributions of microvilli at the cell membrane in various seeding conditions are shown in SI-2, whereas SI-3 displays transmission optical microscopy images of cells incubated with γ-$Fe_2O_3$ NPs. The stability of the coated and bare particles in cellular media is studied in SI-4.

The release amounts and release rates of ferric ion $Fe^{3+}$ at neutral and acidic pH for the Massart dispersions are estimated in SI-5. In SI-6, MTT toxicity assays insure that the NPs used in this study do not present acute short-time toxicity, and in SI-7 the UV-Vis absorption versus wavelength is shown for comparison. This information is available free of charge *via* the Internet at xxx.